\documentclass{article}
\usepackage{amssymb}

\usepackage{amsmath}

\input{tcilatex}

\begin{document}

\title{Alternative descriptions and bipartite compound quantum systems}
\author{G. Scolarici\thanks{%
e-mail: scolarici@le.infn.it} and L. Solombrino\thanks{%
e-mail: solombrino@le.infn.it} \\
Dipartimento di Fisica dell'Universit\`{a} del Salento and\\
INFN, Sezione di Lecce, I-73100 Lecce, Italy }
\maketitle

\begin{abstract}
We analyze some features of alternative Hermitian and
quasi-Hermitian quantum descriptions of simple and bipartite
compound systems. We show that alternative descriptions of two
interacting subsystems are possible if and only if the metric
operator of the compound system can be obtained as tensor product of
positive operators on component spaces. Some examples also show that
such property could be strictly connected with symmetry properties
of the non-Hermitian Hamiltonian.
\end{abstract}

\section{\protect\bigskip Introduction}

In the past years, since a conjecture due to Bender and Boettcher \cite{ben1}%
, a growing interest has been witnessed in $PT$-symmetric non-Hermitian
Hamiltonians with real spectra \cite{proc}, \cite{ben2}. Today, it is well
known that $PT$-symmetry actually constitutes a concrete, physically
relevant realization of $\eta $-pseudo-Hermiticity property \cite{mosta12}, %
\cite{mosta2} defined by relation

\begin{equation}
\eta H\eta ^{-1}=H^{\dagger },  \label{psecond}
\end{equation}%
with $\eta $ Hermitian and invertible.

In the context of the pseudo-Hermitian quantum theory (PHQM) a
relevant role is played by the quasi-Hermitian Hamiltonians, i. e.,
those pseudo-Hermitian Hamiltonians admitting a positive definite
inner product invariant under their dynamics \cite{geyer}. Although
non-Hermitian, these Hamiltonians turn out to be sufficiently close
to those of conventional quantum mechanics
(QM) (in particular they are necessarily diagonalizable with real spectrum %
\cite{mosta2}, \cite{noi1}) and therefore a standard quantum language is
allowed to describe the predicted results.

Actually, a complete mathematical equivalence between PHQM and QM can be
proven, at least if one considers simple quantum systems, since a unitary
mapping exists which connects the corresponding Hilbert spaces \cite{mosta03}%
, \cite{mo1}, \cite{mo2}; more directly, such equivalence can also be proven
by showing the equivalence of the spectra associated with the two systems %
\cite{jo} (we recall however that the physical interpretation of PHQM is
still controversial, as the living debate on the quantum brachistochrone
proves \cite{bra1}, \cite{mo2}, \cite{bra2}). Finally, let us notice that
these theories can be considered alternative Hamiltonian descriptions for
quantum systems (see for instance \cite{dub} and references therein).

Now, a remarkable picture of Hermitian and quasi-Hermitian dynamics,

\begin{equation}
i\frac{d}{dt}|\psi \rangle =H|\psi \rangle  \label{scr}
\end{equation}%
is the existence of a (possibly infinite) set of dynamically
invariant $\eta $-inner product characterized by positive operators
$\eta $ \cite{mosta12} and defined by

\begin{equation}
h_{\eta }(.,.)=h_{1}(.,\eta .),  \label{altpseinnpro}
\end{equation}%
where $h_{1}(.,.)$ denotes the standard or fiducial inner product in the
Hilbert space $\mathcal{H}$. Hence, the possibility of alternative quantum
descriptions naturally arises in this context so that we will study in depth
this topics both for Hermitian and for quasi-Hermitian Hamiltonians. In
particular, in Sec. 2, we will show that the expectation value of the energy
observable strongly depends on the alternative inner product, and that to
different alternative descriptions associated with the same (non pure)
density state, correspond different values of von Neumann entropy. These
results can contribute, in our opinion, to clarify the real meaning of the
above mentioned equivalence between PHQM and QM.

Moreover, since, of course, any physically meaningful theory must be
able to describe compound systems, we will extend PHQM to include
bipartite quantum systems, in order to verify if the equivalence
between PHQM and QM (which, as we said above, has been already
stated in literature only for simple systems) can be proven also at
the level of compound systems. We recall that the problem to analyze
to what extent alternative quantum descriptions survive when one
considers compound systems and interactions among them was already
raised on in Ref. \cite{morandi} with respect to Hermitian dynamics.
We will show in this paper that alternative descriptions play a
crucial role in this respect; indeed, in Sec. 3, we prove a
necessary and sufficient condition (propositions 2 and 3) which
ensures such equivalence. In particular, we prove that
quasi-Hermitian descriptions for bipartite compound quantum systems
are permitted if and only if the positive operator which
characterizes the alternative inner product can be written as the
tensor product of two positive operators on the component spaces.
(To avoid technicalities, we limit ourselves to consider here
finite-dimensional bipartite systems.)\ As a consequence, severe
restrictions arise about the equivalence between PHQM and QM for
compound systems.

Then, these general results are illustrated by some examples in Sec. 4,
where also reduced density matrices via partial traces are introduced; in
particular, the example in Subsec.4.3 shows a physical situation where
alternative descriptions of two interacting subsystems associated with a
quasi-Hermitian Hamiltonian are forbidden. Some concluding remarks are drawn
in the last section.

\section{Alternative quasi-Hermitian descriptions}

We begin by discussing alternative descriptions for quantum systems in the
case of quasi-Hermitian dynamics.

The spectral representation of a $\eta $-quasi-Hermitian Hamiltonian
operator $H$ with a nondegenerate spectrum in terms of its biorthonormal
eigenbasis, $\{|\psi _{n}\rangle ,|\phi _{n}\rangle \}$, reads \cite{mosta2}

\begin{equation}
H=\sum_{n}E_{n}|\psi _{n}\rangle \langle \phi _{n}|,\text{ \ \ \ }E_{n}\in
\mathbb{R}.  \label{sp}
\end{equation}%
Furthermore,

\begin{equation}
\eta =\sum_{n}|\phi _{n}\rangle \langle \phi _{n}|  \label{eta}
\end{equation}%
and%
\begin{equation*}
\eta |\psi _{n}\rangle =|\phi _{n}\rangle .
\end{equation*}

By using the inner product (\ref{altpseinnpro}) and the spectral
representation (\ref{sp}), the expectation value of $H$ in a (normalized)
state,

\begin{equation*}
\frac{1}{\sqrt{\langle \psi |\eta |\psi \rangle }}|\psi \rangle =\frac{1}{%
\sqrt{\langle \psi |\eta |\psi \rangle }}\sum_{n}|\psi _{n}\rangle \langle
\phi _{n}|\psi \rangle ,
\end{equation*}%
can be computed:

\begin{eqnarray}
h_{\eta }(\psi ,H\psi ) &\doteqdot &\sum_{n}E_{n}p(E_{n})=\sum_{n}E_{n}\frac{%
\langle \psi |\eta |\psi _{n}\rangle \langle \psi _{n}|\eta |\psi \rangle }{%
\langle \psi |\eta |\psi \rangle }=  \label{expec} \\
&=&\sum_{n}E_{n}\frac{\langle \psi _{n}|\eta |\psi \rangle \langle \psi
|\eta |\psi _{n}\rangle }{\langle \psi |\eta |\psi \rangle }=\sum_{n}E_{n}%
\frac{\langle \phi _{n}|\psi \rangle \langle \psi |\eta |\psi _{n}\rangle }{%
\langle \psi |\eta |\psi \rangle }=  \notag \\
&=&\sum_{n}\frac{\langle \phi _{n}|\psi \rangle \langle \psi |\eta H|\psi
_{n}\rangle }{\langle \psi |\eta |\psi \rangle }=\mathrm{Tr}\left( \frac{%
|\psi \rangle \langle \psi |\eta }{\langle \psi |\eta |\psi \rangle }%
H\right) =\mathrm{Tr}\tilde{\rho}H  \notag
\end{eqnarray}%
where
\begin{equation*}
\tilde{\rho}=\frac{|\psi \rangle \langle \psi |\eta }{\langle \psi |\eta
|\psi \rangle }.
\end{equation*}%
More generally, if $\rho $ denotes a generic (Hermitian, positive definite)
density matrix ($\mathrm{Tr}\rho =1$), we can associate with it a
generalized density matrix $\tilde{\rho}$ by means of the one-to-one mapping
in the following way \cite{bologna}, \cite{mo2}:

\begin{equation}
\tilde{\rho}=\frac{\rho \eta }{\mathrm{Tr}\rho \eta }  \label{Rotilda}
\end{equation}%
and obtain
\begin{equation*}
\langle H\rangle _{\eta }=\mathrm{Tr}\tilde{\rho}H.
\end{equation*}

The dynamics of $\tilde{\rho}$ is ruled at infinitesimal level by the
Liouville-von Neumann equation \cite{bologna}
\begin{equation}
\frac{d}{dt}\tilde{\rho}\left( t\right) =-i[H,\tilde{\rho}].
\label{liovquasi}
\end{equation}%
We note explicitly that the mapping (\ref{Rotilda}) does not change the rank
of the density matrices \cite{horn},

\begin{equation}
\mathrm{rank}\tilde{\rho}=\mathrm{rank}\rho .  \label{rank}
\end{equation}%
Of course, in such scheme, to different metric operators $\eta $ and $\eta
^{\prime }$, both fulfilling relation (\ref{psecond}), correspond
''alternative'' descriptions of the same quantum system.

It is clear that changing the inner product corresponds to different
expectation values of $H$ on the same state $|\psi \rangle $. In fact, from
Eq. (\ref{expec}), we obtain in general,

\begin{equation}
\langle H\rangle _{\eta }=\mathrm{Tr}\left( \frac{\rho \eta }{\mathrm{Tr}%
\rho \eta }H\right) \neq \mathrm{Tr}\left( \frac{\rho \eta ^{\prime }}{%
\mathrm{Tr}\rho \eta ^{\prime }}H\right) =\langle H\rangle _{\eta ^{\prime
}}.  \label{diffexpect}
\end{equation}

Moreover, denoting with $S(\widetilde{\rho })$ the von Neumann entropy
associated with a density matrix $\widetilde{\rho }$, as a consequence of
Eq. (\ref{diffexpect}) we immediately get

\begin{equation*}
S(\widetilde{\rho })=-\mathrm{Tr}(\widetilde{\rho }\log \widetilde{\rho }%
)\neq -\mathrm{Tr}(\widetilde{\rho }^{\prime }\log \widetilde{\rho }^{\prime
})=S(\widetilde{\rho }^{\prime }),
\end{equation*}%
i. e., the von Neumann entropy depends on the alternative inner product.
Then, we can conclude that the entropy and the expectation value of
quasi-Hermitian observables strongly depend on the alternative description
we consider on $\mathcal{H}$.

Note that if $\eta H\eta ^{-1}=H^{\dagger }$ and $\eta ^{\prime }H\eta
^{\prime -1}=H^{\dagger }$ with $\eta ^{\prime }\neq \eta $, by performing
in the Hilbert space $\mathcal{H}$ the linear transformation induced by $%
\eta ^{\frac{1}{2}}$, we get

\begin{equation}
H\rightarrow H^{\prime }=\eta ^{\frac{1}{2}}H\eta ^{-\frac{1}{2}}=H^{\prime
\dagger },  \label{A}
\end{equation}%
while the metric operators transform by congruence \cite{andrea}:

\begin{equation}
\eta \rightarrow \eta ^{-\frac{1}{2}}\eta (\eta )^{\dagger -\frac{1}{2}%
}=\eta ^{-\frac{1}{2}}\eta \eta ^{-\frac{1}{2}}=\mathbf{1,}  \label{B}
\end{equation}

\begin{equation*}
\eta ^{\prime }\rightarrow \eta ^{-\frac{1}{2}}\eta ^{\prime }\eta ^{-\frac{1%
}{2}}
\end{equation*}%
and the $\eta ^{\prime }$-quasi-Hermiticity condition of $H$ implies

\begin{equation*}
\lbrack H^{\prime },\eta ^{-\frac{1}{2}}\eta ^{\prime }\eta ^{-\frac{1}{2}%
}]=0,
\end{equation*}%
i. e., $\eta ^{-\frac{1}{2}}\eta ^{\prime }\eta ^{-\frac{1}{2}}$ belongs to
the commutant of $H^{\prime }$ and represents the operator which connect an
alternative inner product invariant under time-translation generated by $%
H^{\prime }$ to the fiducial scalar product (see also \cite{mm}). Then, in
the space where $H$ becomes Hermitian alternative inner products can be
obtained by means of positive definite operators in the commutant of $%
H^{\prime }$ (coming back, the mapping (\ref{A}), (\ref{B}), can also be
useful to compute the full set of $\eta $ operators fulfilling the
quasi-Hermiticity condition). Moreover, the same calculations show that all
the statements for quasi-Hermitian Hamiltonians also hold, with minimal and
obvious changes, for the Hermitian ones.

\section{Quasi-Hermitian bipartite quantum systems}

Let us consider a compound bipartite quantum system which dynamics is
described in a Hilbert space $\mathcal{H}^{nm}$ of dimension $nm$ by an
evolution operator $U$ such that

\begin{equation*}
U^{\dagger }\eta U=\eta ,
\end{equation*}%
where

\begin{equation*}
U(t)=e^{-iHt}
\end{equation*}%
and the time-independent Hamiltonian $H$ satisfies,

\begin{equation*}
\eta H\eta ^{-1}=H^{\dagger },\text{ \ \ }\eta >0.
\end{equation*}%
Then, the alternative Hermitian structure, $h_{\eta }(.,.)=h_{1}(.,\eta .)$,
is invariant under the dynamics generated by $H$.

Now, a natural question arises: Is it possible a proper quantum mechanical
description of such quasi-Hermitian compound quantum systems in terms of
their corresponding component systems?

In order to answer to this question, as a preliminary step, we put the
following proposition which gives a necessary condition for an operator $%
\eta $ of dimension $nm$ to be written as the tensor product of two
operators of dimension $n$ and $m$ respectively: $\eta =\xi \otimes \zeta $.

\textbf{Proposition 1. }\textit{A positive Hermitian operator }$\eta $%
\textit{\ with eigenvalues }$\{\eta _{ij}:i=1,2,...,n;
j=1,2,...,m \}$ \textit{acting on the complex vector space }$\mathcal{H}%
^{nm}$ \textit{of dimension }$nm$\textit{,\ can be decomposed as }$\eta =\xi
\otimes \zeta $\textit{\ where }$\xi $\textit{\ and }$\zeta $\textit{\
represent positive Hermitian operators acting on }$\mathcal{H}^{n}$\textit{\
and }$\mathcal{H}^{m}$\textit{\ with eigenvalues }$\{\xi _{i}:$ $%
i=1,2,...,n\}$\textit{\ and }$\{\zeta _{j}:$ $j=1,2,...,m\}$\textit{\
respectively, only if the following }$n^{2}$ \textit{and} $m^{2}$\textit{\
constraints are satisfied: }$\frac{\eta _{ij}}{\xi _{i}}=\frac{\eta
_{i^{\prime }j}}{\xi _{i^{\prime }}}$, $i,$\textit{\ }$i^{\prime }=1,2,...,n$
\textit{and\ }$\frac{\eta _{ij}}{\zeta _{j}}=\frac{\eta _{ij^{\prime }}}{%
\zeta _{j^{\prime }}}$\textit{, }$j,$\textit{\ }$j^{\prime }=1,2,...,m$%
\textit{.}

\textbf{Proof. }Let us suppose $\eta =\xi \otimes \zeta $. Then,
diagonalizing $\xi $ and $\zeta $ (and suitably ordering the spectra) we get

\begin{equation*}
\xi _{i}\zeta _{j}=\eta _{ij}
\end{equation*}%
where $i=1,2,...,n$,\textit{\ }$j=1,2,...,m.$ From the invertibility of $\xi
$ and $\zeta $ we immediately get

\begin{equation*}
\frac{\eta _{ij}}{\xi _{i}}=\frac{\eta _{i^{\prime }j}}{\xi _{i^{\prime }}}%
,i,\mathit{\ }i^{\prime }=1,2,...,n\mathit{\ \ \ \ \ \ }\text{or \ \ \ \ }%
\frac{\eta _{ij}}{\zeta _{j}}=\frac{\eta _{ij^{\prime }}}{\zeta _{j^{\prime
}}},\mathit{\ }j,\mathit{\ }j^{\prime }=1,2,...,m\mathit{.}
\end{equation*}%
$\square $

On the other hand, a sufficient condition in order that a given $\eta $
operator satisfying the previous constraints can be written as $\eta =\xi
\otimes \zeta $, is that it must be diagonalizable by means of a unitary
transformation of the form $U_{1}\otimes U_{2}$ where $U_{1}\in U(n,\mathbf{C%
})$ and $U_{2}\in U(m,\mathbf{C})$.

Now, we denote with

\begin{equation*}
U(nm,\mathbf{C},h_{\eta }),
\end{equation*}%
the group which preserve the alternative Hermitian structure $h_{\eta }$.

Having in mind component systems, and recalling that in standard QM for any $%
U_{1}\in U(n,\mathbf{C})$ and $U_{2}\in U(m,\mathbf{C})$

\begin{equation*}
U_{1}\otimes U_{2}\in U(nm,\mathbf{C}),
\end{equation*}%
the following proposition gives a necessary and sufficient condition for the
Hilbert spaces $\mathcal{H}^{n}$ and $\mathcal{H}^{m}$ associated with the
component systems to be provided of suitable alternative Hermitian
structures $h_{\xi }$ and $h_{\zeta }$ such that for any $U_{\xi }\in U(n,%
\mathbf{C},h_{\xi })$ and $U_{\zeta }\in U(m,\mathbf{C},h_{\zeta })$

\begin{equation*}
U_{\xi }\otimes U_{\zeta }\in U(nm,\mathbf{C},h_{\eta }).
\end{equation*}

\textbf{Proposition 2.} \textit{For any }$U_{\xi }\in U(n,\mathbf{C},h_{\xi
})$ \textit{and} $U_{\zeta }\in U(m,\mathbf{C},h_{\zeta })$, \textit{the
group} $U(nm,\mathbf{C},h_{\eta })$ \textit{contains the transformations}%
\begin{equation*}
U_{\xi }\otimes U_{\zeta }
\end{equation*}%
\textit{\ if and only if }%
\begin{equation*}
\eta =\xi \otimes \zeta \mathit{.}
\end{equation*}%
Denoting with $\mathfrak{H}_{\xi }^{n}$, $%
\mathfrak{H}_{\zeta }^{m}$ and $\mathfrak{H}_{\eta }^{nm}$ the group
algebras associated with $U(n,\mathbf{C},h_{\xi })$,
$U(m,\mathbf{C},h_{\zeta })$ and $U(nm,\mathbf{C},h_{\eta })$
respectively, proposition 2 can be equivalently restated in the
following form:

\textbf{Proposition 3.} \textit{The set} $\mathfrak{H}_{\eta }^{nm}$ \textit{%
of} $\eta $\textit{-quasi-Hermitian matrices of dimension }$nm$ \textit{%
contain as its subset}

\begin{equation*}
\emph{K}^{nm}=\{O_{\xi }\otimes O_{\zeta }\text{ }|\text{ }O_{\xi
}\in \mathfrak{H}%
_{\xi }^{n},O_{\zeta }\in \text{ }\mathfrak{H}_{\zeta }^{m}\},
\end{equation*}%
\textit{where} $\mathfrak{H}_{\xi }^{n}$ \textit{represent the set of} $\xi $%
\textit{-quasi-Hermitian matrices} \textit{of dimension }$n$ \textit{and} $%
\mathfrak{H}_{\zeta }^{m}$ \textit{represent the set of }$\zeta $\textit{%
-quasi-Hermitian matrices of dimension }$m$\textit{, if and only if }%
\begin{equation*}
\eta =\xi \otimes \zeta \mathit{.}
\end{equation*}

\textbf{Proof. }Let us suppose $\eta =\xi \otimes \zeta $. Then,
trivially, the set $\emph{K}^{nm}$ is constituted by $\eta
$-quasi-Hermitian matrices, hence, $\emph{K}^{nm}\cap
\mathfrak{H}_{\eta }^{nm}\equiv \emph{K}^{nm}$. Conversely, let us
suppose $\eta \neq \xi \otimes \zeta $ for any positive operator
$\xi $ and $\zeta $. The set $\emph{K}^{nm}$ is obviously an
irreducible set of $\xi \otimes \zeta $-quasi-Hermitian matrices;
then, by a known result in literature (see Ref.$\cite{geyer}$), the
metric is unique (up to a normalization factor), hence, it coincides with $%
\xi \otimes \zeta $ . Then, the set $\emph{K}^{nm}$ contain some matrices
that cannot be $\eta $-quasi-Hermitians, hence, $\emph{K}^{nm}\cap \mathfrak{%
H}_{\eta }^{nm}\supset \emph{K}^{nm}$. $\square $

A direct consequence of proposition 3 is that the tensor product of two
observables in ($\mathcal{H}^{n}$, $h_{\xi }$) and ($\mathcal{H}^{m}$, $%
h_{\zeta }$), is certainly an observable in ($\mathcal{H}^{nm}$, $h_{\eta }$%
), if and only if $\eta =\xi \otimes \zeta $. In particular, let us assume, $%
\eta \neq \xi \otimes \zeta $. Then, some elements of the set $\{O_{\xi
}\otimes \mathbf{1}^{m},\mathbf{1}^{n}\otimes O_{\zeta }\}$, cannot be
observable. In fact, let us suppose

\begin{equation*}
\eta \left( O_{\xi }\otimes \mathbf{1}^{m}\right) \eta ^{-1}=O_{\xi
}^{\dagger }\otimes \mathbf{1}^{m}
\end{equation*}%
and

\begin{equation*}
\eta \left( \mathbf{1}^{n}\otimes O_{\zeta }\right) \eta ^{-1}=\mathbf{1}%
^{n}\otimes O_{\zeta }^{\dagger }
\end{equation*}%
for any $O_{\xi }$ and $O_{\zeta }$. From the commutativity,

\begin{equation*}
\lbrack O_{\xi }\otimes \mathbf{1}^{m},\mathbf{1}^{n}\otimes O_{\zeta }]=0,
\end{equation*}%
we immediately get

\begin{equation*}
\eta \left( O_{\xi }\otimes O_{\zeta }\right) \eta ^{-1}=O_{\xi }^{\dagger
}\otimes O_{\zeta }^{\dagger }
\end{equation*}%
whereas, according with the above hypothesis,%
\begin{equation*}
(\xi \otimes \zeta )\left( O_{\xi }\otimes O_{\zeta }\right) (\xi \otimes
\zeta )^{-1}=O_{\xi }^{\dagger }\otimes O_{\zeta }^{\dagger }.
\end{equation*}%
From the irreducibility of the set $\{O_{\xi }\otimes O_{\zeta }\}$, (see
proposition 3) the thesis follows at once.

Then we can conclude that any $\eta $-quasi-Hermitian compound quantum
system admits a proper quantum mechanical description in terms of component
systems if and only if $\eta =\xi \otimes \zeta $.

Note, of course, that if an operator $\eta $ admits a decomposition $\eta
=\xi \otimes \zeta $ , such decomposition is not unique. In fact, we can for
instance change the operators $\xi $ and $\zeta $ by multiplying them by
(positive) factors $r$ and $\frac{1}{r}$ respectively.

\textbf{Remark.} Clearly if $\eta =\xi \otimes \zeta $ the peculiarity of a
state, associated with a compound system, to be entangled or not does not
depend on the alternative structures $h_{\xi }$ and $h_{\zeta }$ on
component spaces. In fact, let us consider

\begin{equation*}
|\beta \rangle =|\chi \rangle \otimes |\omega \rangle \in \mathcal{H}^{nm}.
\end{equation*}%
Then,

\begin{equation*}
|\beta ^{\prime }\rangle =S|\beta \rangle
\end{equation*}%
is entangled if and only if $S\in GL(nm,\mathbb{C})$ and $S\neq S_{1}\otimes
S_{2}$ for any $S_{1}\in GL(n,\mathbb{C})$ and $S_{2}\in GL(m,\mathbb{C})$.
Hence, the (alternative) Hermitian structure does not play here any role.

\section{Examples}

We illustrate the general results in the previous sections by means of some
examples. In order to do that, we first introduce the reduced density
matrices for bipartite quasi-Hermitian systems via partial trace operation.

Let be given a $\eta $-quasi-Hermitian Hamiltonian associated with a
bipartite system,

\begin{equation}
H=H_{A}\otimes \mathbf{1}^{m}+\mathbf{1}^{n}\otimes H_{B}+V_{int}
\label{Hbiptrite}
\end{equation}%
where we assume for the sake of simplicity that

\begin{equation}
H_{A}=\sum_{i=1}^{n}a_{i}|\psi _{i}\rangle \langle \phi _{i}|,\ \ \ a_{i}\in
\mathbb{R}  \label{Ha}
\end{equation}%
and

\begin{equation}
H_{B}=\sum_{j=1}^{m}b_{j}|\Psi _{j}\rangle \langle \Phi _{j}|,\ \ \ b_{j}\in
\mathbb{R}  \label{Hb}
\end{equation}%
are quasi-Hermitians with a nondegenerate spectrum. Then, the more general $%
\xi $ and $\zeta $ operators satisfying the quasi-Hermiticity conditions, $%
\xi H_{A}\xi ^{-1}=H_{A}^{\dagger }$ and $\zeta H_{B}\zeta
^{-1}=H_{B}^{\dagger }$, are respectively given by

\begin{equation}
\xi =\sum_{i=1}^{n}r_{i}|\phi _{i}\rangle \langle \phi _{i}|,\ \ r_{i}>0
\label{csi}
\end{equation}%
and

\begin{equation}
\zeta =\sum_{j=1}^{m}s_{j}|\Phi _{j}\rangle \langle \Phi _{j}|,\ \ s_{j}>0.
\label{zita}
\end{equation}

Then, any state $\frac{|\alpha \rangle }{\langle \alpha |\eta |\alpha
\rangle }$ in the space $\mathcal{H}^{nm}$ provided with Hermitian structure
$h_{\eta }$, can be decomposed on the biorthonormal basis $\left\{ |\psi
_{i}\rangle \otimes |\Psi _{j}\rangle ,|\phi _{i}\rangle \otimes |\Phi
_{j}\rangle \right\} $:

\begin{eqnarray*}
&&\frac{1}{\langle \alpha |\eta |\alpha \rangle }|\alpha \rangle \\
&=&\left( \sum_{i,j}|\psi _{i}\rangle \otimes |\Psi _{j}\rangle (\langle
\Phi _{j}|\otimes \langle \phi _{i}|)\right) \frac{1}{\langle \alpha |\eta
|\alpha \rangle }|\alpha \rangle ,
\end{eqnarray*}%
and the associated rank-one density matrix reads

\begin{equation*}
\text{\ }\widetilde{\rho }^{AB}=\frac{|\alpha \rangle \langle \alpha |\eta }{%
\langle \alpha |\eta |\alpha \rangle }.
\end{equation*}%
Then, we immediately get

\begin{equation}
\widetilde{\rho }_{A}=\mathrm{Tr}_{B}\widetilde{\rho }^{AB}=\sum_{j=1}^{m}%
\langle \Phi _{j}|\widetilde{\rho }^{AB}|\Psi _{j}\rangle  \label{par1}
\end{equation}%
and

\begin{equation}
\widetilde{\rho }_{B}=\mathrm{Tr}_{A}\widetilde{\rho }^{AB}=\sum_{i=1}^{n}%
\langle \phi _{i}|\widetilde{\rho }^{AB}|\psi _{i}\rangle .  \label{par2}
\end{equation}%
Moreover, being $\eta =\xi \otimes \zeta $, we obtain

\begin{equation}
\widetilde{\rho }_{A}=\mathrm{Tr}_{B}\widetilde{\rho }^{AB}=\frac{\rho
^{A}\xi }{\mathrm{Tr}\rho ^{A}\xi }\text{ \ \ and \ \ }\widetilde{\rho }_{B}=%
\mathrm{Tr}_{A}\widetilde{\rho }^{AB}=\frac{\rho ^{B}\zeta }{\mathrm{Tr}\rho
^{B}\zeta },  \label{usualpart}
\end{equation}%
where $\rho ^{A}$ and $\rho ^{B}$ denote the partial traces associated with
the state $\frac{|\alpha \rangle \langle \alpha |}{\langle \alpha |\alpha
\rangle }$ in the fiducial (standard) description.

Now, we will consider some examples. In the first one, a dynamics generated
by an Hermitian Hamiltonian associated with a composite system on a four
dimensional Hilbert space is described in terms of two alternative invariant
inner products. In the second one, alternative descriptions of a $PT$%
-symmetric quasi-Hermitian dynamics recently introduced in literature are
considered. In the third one, alternative descriptions of a quasi-Hermitian
not $PT$-symmetric dynamics are considered.

\subsection{An Hermitian Hamiltonian}

Because of the physical relevance of two qubit quantum gates, we shall now
consider alternative descriptions for an optimal entanglement generation
recently introduced in literature \cite{sudarshan}. The system we consider
is composed of two qubits $A$ and $B$, hence $\mathcal{H}\equiv \mathbb{C}%
^{4}$.

The Hamiltonian and the evolution operator of the overall system are

\begin{eqnarray}
H &=&\sigma _{3}^{A}\otimes \mathbf{1}^{B}+\mathbf{1}^{A}\otimes \sigma
_{3}^{B}+V_{int}=  \label{H1} \\
&&\left(
\begin{array}{cccc}
1 & 0 & 0 & 0 \\
0 & -1 & 0 & 0 \\
0 & 0 & -1 & 0 \\
0 & 0 & 0 & 1%
\end{array}%
\right) ,  \notag
\end{eqnarray}

\begin{eqnarray}
U &=&\cos t\mathbf{1}^{A}\otimes \mathbf{1}^{B}-i\sin t\sigma
_{3}^{A}\otimes \sigma _{3}^{B}=  \label{U1} \\
&&\left(
\begin{array}{cccc}
e^{-it} & 0 & 0 & 0 \\
0 & e^{it} & 0 & 0 \\
0 & 0 & e^{it} & 0 \\
0 & 0 & 0 & e^{-it}%
\end{array}%
\right) .  \notag
\end{eqnarray}

Let the initial state be

\begin{equation*}
\rho ^{A}(0)\otimes \rho ^{B}(0)=\frac{1}{2}\left(
\begin{array}{cc}
1 & 1 \\
1 & 1%
\end{array}%
\right) \otimes \frac{1}{2}\left(
\begin{array}{cc}
1 & -i \\
i & 1%
\end{array}%
\right) .
\end{equation*}

At time $t$ we get

\begin{equation*}
\rho ^{AB}(t)=U(t)\rho ^{AB}(0)U(t)^{\dagger }
\end{equation*}%
and we obtain by partial traces the final states

\begin{eqnarray*}
\rho ^{A}(t) &=&\mathrm{Tr}_{B}\rho ^{AB}=\frac{1}{2}\left(
\begin{array}{cc}
1 & \cos 2t \\
\cos 2t & 1%
\end{array}%
\right) , \\
\rho ^{B}(t) &=&\mathrm{Tr}_{A}\rho ^{AB}=\frac{1}{2}\left(
\begin{array}{cc}
1 & -i\cos 2t \\
i\cos 2t & 1%
\end{array}%
\right) .
\end{eqnarray*}%
We stress that at the time $t_{bell}=\pi /4$, the overall state $\rho
^{AB}(t=t_{bell})$ is equivalent to a Bell state $\frac{1}{\sqrt{2}}%
(|00\rangle +|11\rangle )$.

The von Neumann entropy gives an entanglement measure:

\begin{eqnarray}
S(\rho ^{A}(t)) &=&-\mathrm{Tr}\text{(}\rho ^{A}\log \rho ^{A}\text{)}=
\label{Sro} \\
&&-(\sin ^{2}t)\log (\sin ^{2}t)-(\cos ^{2}t)\log (\cos ^{2}t).  \notag
\end{eqnarray}%
In particular, $S(\rho ^{A}(t))=0$ when the state $\rho ^{AB}(t)$ becomes
separable and this happens when the purity \cite{sudarshan} of both the
reduced density matrices,

\begin{eqnarray}
P_{\rho ^{A}(t)} &=&\mathrm{Tr}\rho ^{A}(t)^{2}=\frac{1}{2}\left( 1+(\cos
2t)^{2}\right) ,  \label{Pro} \\
P_{\rho ^{B}(t)} &=&\mathrm{Tr}\rho ^{B}(t)^{2}=\frac{1}{2}\left( 1+(\cos
2t)^{2}\right) ,  \notag
\end{eqnarray}%
becomes $1$, that is when $t=\frac{k\pi }{2}$, $k\in \mathbb{N}$.

Now, let us consider in the Hilbert space associated with the compound
system the most general (in the sense of proposition 2) alternative scalar
product which is connected with the fiducial one by means of the (positive)
operator,

\begin{equation}
\eta =\xi \otimes \zeta =\left(
\begin{array}{cc}
\xi _{1} & 0 \\
0 & \xi _{2}%
\end{array}%
\right) \otimes \left(
\begin{array}{cc}
\zeta _{1} & 0 \\
0 & \zeta _{2}%
\end{array}%
\right) .  \label{esaltmetr}
\end{equation}

The Hermitian structures $h_{\eta }$ and $h_{\xi }$, $h_{\zeta }$ are well
defined alternative inner products for composite and component systems
respectively. In fact,

\begin{equation*}
\lbrack H,\eta ]=0,\text{ \ \ \ }U^{\dagger }\eta U=\eta ,
\end{equation*}

\begin{equation*}
\lbrack \sigma _{3}^{A},\xi ]=0,\text{ \ }[\sigma _{3}^{B},\zeta ]=0,
\end{equation*}%
hence, $H$, $\sigma _{3}^{A}$ and $\sigma _{3}^{B}$ are also Hermitian with
respect to $h_{\eta }$, $h_{\xi }$ and $h_{\zeta }$, respectively.

The initial density matrix in the description associated with $h_{\eta }$
reads

\begin{equation*}
\widetilde{\rho }^{AB}(0)=\frac{\left( \rho ^{A}(0)\otimes \rho
^{B}(0)\right) \eta }{\mathrm{Tr}\left( \rho ^{A}(0)\otimes \rho
^{B}(0)\right) \eta }.
\end{equation*}

At time $t$ we get

\begin{equation*}
\widetilde{\rho }^{AB}(t)=U(t)\widetilde{\rho }^{AB}(0)U(t)^{\dagger }
\end{equation*}%
and from Eq. (\ref{usualpart}), the reduced density matrices can be computed:

\begin{eqnarray*}
\widetilde{\rho }^{A}(t) &=&\mathrm{Tr}_{B}\widetilde{\rho }^{AB}= \\
&&\frac{1}{\xi _{1}+\xi _{2}}\left(
\begin{array}{cc}
\xi _{1} & \xi _{2}\cos 2t \\
\xi _{1}\cos 2t & \xi _{2}%
\end{array}%
\right) ,
\end{eqnarray*}

\begin{eqnarray*}
\widetilde{\rho }^{B}(t) &=&\mathrm{Tr}_{A}\widetilde{\rho }^{AB}= \\
&&\frac{1}{\zeta _{1}+\zeta _{2}}\left(
\begin{array}{cc}
\zeta _{1} & -i\zeta _{2}\cos 2t \\
i\zeta _{1}\cos 2t & \zeta _{2}%
\end{array}%
\right) .
\end{eqnarray*}%
Note that $\widetilde{\rho }^{A}(t)$ and $\widetilde{\rho }^{B}(t)$ are $\xi
$- and $\zeta $-quasi-Hermitian respectively.

The eigenvalues of $\widetilde{\rho }^{A}(t)$ are:

\begin{equation*}
r_{\pm }=\frac{1}{2}\left( 1\pm \frac{\sqrt{\xi _{1}^{2}+\xi _{2}^{2}+2\xi
_{1}\xi _{2}\cos 4t}}{\xi _{1}+\xi _{2}}\right) ,
\end{equation*}%
hence, its von Neumann entropy reads now

\begin{eqnarray}
S(\widetilde{\rho }^{A}(t)) &=&-\mathrm{Tr}\text{(}\widetilde{\rho }^{A}\log
\widetilde{\rho }^{A}\text{)}=  \label{Srotilda} \\
&&-r_{+}\log r_{+}-r_{-}\log r_{-}.  \notag
\end{eqnarray}%
It is then evident that, the entropy depends on the alternative scalar
product, in fact, from Eqs. (\ref{Sro}), (\ref{Srotilda}), we immediately get

\begin{equation*}
S(\rho ^{A}(t))\neq S(\widetilde{\rho }^{A}(t)).
\end{equation*}%
Then, the entanglement measure strongly depends on the alternative Hermitian
structure.

It is worthwhile to note however that, $S(\widetilde{\rho }^{A}(t))=0$ when
the purity of both the reduced density matrices

\begin{eqnarray}
P_{\widetilde{\rho }^{A}(t)} &=&\mathrm{Tr}\widetilde{\rho }^{A}(t)^{2}=%
\frac{1}{2}\left( 1+\frac{\xi _{1}^{2}+\xi _{2}^{2}+2\xi _{1}\xi _{2}\cos 4t%
}{\left( \xi _{1}+\xi _{2}\right) ^{2}}\right) ,  \label{Protilda} \\
P_{\widetilde{\rho }^{B}(t)} &=&\mathrm{Tr}\widetilde{\rho }^{B}(t)^{2}=%
\frac{1}{2}\left( 1+\frac{\zeta _{1}^{2}+\zeta _{2}^{2}+2\zeta _{1}\zeta
_{2}\cos 4t}{\left( \zeta _{1}+\zeta _{2}\right) ^{2}}\right) ,  \notag
\end{eqnarray}%
becomes $1$ and this happens again when $t=\frac{k\pi }{2}$, $k\in \mathbb{N}
$.

From Eqs. (\ref{Pro}), (\ref{Protilda}) we conclude that whereas the entropy
depends on the alternative inner product, the peculiarity of a state to be
entangled or not does not depend on the alternative description (see also
Eq. (\ref{rank})) as we can expect from general considerations (see the
remark in section 3).

\subsection{A $PT$-symmetric Hamiltonian}

Now, we discuss a recently introduced coupling between two qubits separately
described by a Hermitian Hamiltonian and by a $PT$-symmetric Hamiltonian
respectively \cite{jones}, in terms of compound and component systems. In
particular, we consider the Hermitian Hamiltonian%
\begin{equation}
H_{A}=\left(
\begin{array}{cc}
1 & 0 \\
0 & 1%
\end{array}%
\right)   \label{J1}
\end{equation}%
and the non-Hermitian  Hamiltonian with real eigenvalues

\begin{equation}
H_{B}=\frac{1}{2}\left(
\begin{array}{cc}
\sqrt{3}+i & 2 \\
2 & \sqrt{3}-i%
\end{array}%
\right) ;  \label{J2}
\end{equation}%
both the above Hamiltonians are $PT$-symmetric, where

\begin{equation*}
P=\sigma _{1}=\left(
\begin{array}{cc}
0 & 1 \\
1 & 0%
\end{array}%
\right) ,\text{ \ \ \ }T=K,
\end{equation*}%
($K$ denotes complex conjugation). Then, we couple them by means of nonzero
elements in the off-diagonal sectors, obtaining so (with a coupling constant
$\epsilon =\frac{1}{2}$):

\begin{eqnarray}
H &=&H_{A}\otimes \mathbf{1}^{B}+\mathbf{1}^{A}\otimes H_{B}+V_{int}=
\label{JF} \\
= &&\frac{1}{2}\left(
\begin{array}{cccc}
2 & 0 & 1 & 0 \\
0 & 2 & 0 & 1 \\
1 & 0 & \sqrt{3}+i & 2 \\
0 & 1 & 2 & \sqrt{3}-i%
\end{array}%
\right) .  \notag
\end{eqnarray}

As we said above, such Hamiltonian is obtained by a suitable choice of
parameters from the one considered in \cite{jones}.

The coupling terms are chosen in such a way that $H$ remains invariant under
the combined parity reflection and time reversal, where $T=K$ and

\begin{equation}
P=\mathbf{1}^{A}\otimes \sigma _{1}=\left(
\begin{array}{cccc}
0 & 1 & 0 & 0 \\
1 & 0 & 0 & 0 \\
0 & 0 & 0 & 1 \\
0 & 0 & 1 & 0%
\end{array}%
\right) .  \label{parita}
\end{equation}

Being $\epsilon =\frac{1}{2}$, the eigenvalues of the combined system are
real \cite{jones} and a positive definite $\eta $ can be written as

\begin{equation}
\eta =\mathbf{1}^{A}\otimes \zeta   \label{Jeta}
\end{equation}%
where  $\zeta H_{B}\zeta ^{-1}=H_{B}^{\dagger }$ ; for instance,

\begin{equation*}
\zeta =\left(
\begin{array}{cc}
2 & -\sqrt{2}-i \\
-\sqrt{2}+i & 2%
\end{array}%
\right) .
\end{equation*}

Clearly, by considering the (infinite) set of positive operators $\zeta $
satisfying the quasi-Hermiticity condition, $\zeta H_{B}\zeta
^{-1}=H_{B}^{\dagger }$, we obtain a set of possible $\eta $ operators on $%
\mathbb{C}^{4}$ of the form, $\eta =\xi \otimes \zeta $, permitting
alternative quasi-Hermitian descriptions of our system in terms of
its component systems. For the sake of brevity we do not compute
here the evolution of the overall system, the reduced density
matrices and their von Neumann entropy as in the previous example.

\subsection{A quasi-Hermitian not $PT$-symmetric Hamiltonian}

Finally, let us consider a quasi-Hermitian Hamiltonan $H$ on $\mathbb{C}^{4}$%
, obtained by taking the direct sum of the Hermitian not $PT$-symmetric
Hamiltonian, $H_{A}=\left(
\begin{array}{cc}
1 & 0 \\
0 & 2%
\end{array}%
\right) $, and the $PT$-symmetric Hamiltonian, $H_{B}$ given in Eq. (\ref{J2}%
):

\begin{equation}
H=\left(
\begin{array}{cc}
H_{A} & 0 \\
0 & H_{B}%
\end{array}%
\right) =\frac{1}{2}\left(
\begin{array}{cccc}
2 & 0 & 0 & 0 \\
0 & 4 & 0 & 0 \\
0 & 0 & \sqrt{3}+i & 2 \\
0 & 0 & 2 & \sqrt{3}-i%
\end{array}%
\right) .  \label{l1}
\end{equation}

A direct computation shows that $H$ is not invariant under the combined
parity reflection given by (\ref{parita}) and time-reversal $T=K$ , but it
is surely quasi-Hermitian since it is diagonalizable with real spectrum.

Let us show that the Hamiltonian (\ref{l1}) cannot admit a positive $\eta $
operator, satisfying the quasi-Hermiticity condition, of the form $\eta =\xi
\otimes \zeta $, were $\xi $ and $\zeta $ are positive operators on $\mathbb{%
C}^{2}$. In fact, writing

\begin{equation*}
\xi =\left(
\begin{array}{cc}
a & z \\
z^{\ast } & b%
\end{array}%
\right) ,\text{ \ \ }a,b\in \mathbb{R},\text{ \ \ }z\in \mathbb{C},
\end{equation*}%
the conditions: $\det \xi >0$ $(\Rightarrow ab>0)$ and $\mathrm{Tr}\xi =a+b>0
$, \ which ensure the positivity of $\xi $, imply that $a,b$ must be
non-zero, positive real numbers.

Let us now consider the Kronecker product

\begin{equation*}
\eta =\xi \otimes \zeta =\left(
\begin{array}{cc}
a\zeta & z\zeta \\
z^{\ast }\zeta & b\zeta%
\end{array}%
\right)
\end{equation*}%
and impose

\begin{equation}
\eta H=H^{\dagger }\eta .  \label{pseudoEx}
\end{equation}%
Eq. (\ref{pseudoEx}) is equivalent to the following matrix equations:

\begin{equation}
\zeta H_{A}=H_{A}\zeta ,  \label{a}
\end{equation}

\begin{equation}
\zeta H_{B}=H_{B}^{\dagger }\zeta ,  \label{b}
\end{equation}

\begin{equation}
z(\zeta H_{B}-H_{A}\zeta )=0.  \label{c}
\end{equation}%
From Eq. (\ref{a}) we immediately obtain that $\zeta $ must have a
diagonal form, but a direct computation shows that no diagonal
$\zeta $ can satisfy Eq. (\ref{b}). Then, in this case no
invertible, Hermitian positive operator $\eta $ exists which
satisfies the condition $\eta =\xi \otimes \zeta $.

Then, we conclude that in this case, quasi-Hermitian descriptions
for subsystems are forbidden.

\section{Concluding remarks}

In this paper we considered some features of alternative
descriptions of simple and compound quantum systems and we have
shown, also by means of examples, that the entanglement measure (von
Neumann entropy) strongly depends on the alternative Hermitian
structure. Moreover, we have analyzed to what extent Hermitian and
quasi-Hermitian quantum descriptions of compound systems survive.

The main result of our paper is that if (and only if) the alternative
Hermitian structure is connected with the fiducial ones by means of a
positive operator $\eta $ such that

\begin{equation*}
\eta =\xi \otimes \zeta ,
\end{equation*}%
the projection on the component spaces can be performed via partial trace
operation.

On the contrary, if

\begin{equation*}
\eta \neq \xi \otimes \zeta
\end{equation*}%
quasi-Hermitian descriptions for the component subsystems cannot be
obtained and the corresponding physical theories are inconsistent
(at least, according with the usual physical interpretation of the
mathematical entities).

These results, as a consequence, pose severe restrictions on PHQM,
in particular with respect to the asserted equivalence between such
theories and standard Quantum Mechanics.

In fact, we observe that whereas the set of alternative inner
product associated with Hermitian Hamiltonians admits a not void
subset of operators of the form $\eta =\xi \otimes \zeta $ (in fact
the identity trivially
belongs to this subset), the example in Subsec.4.3 shows that when $\eta $%
-quasi-Hermitian Hamiltonians are considered the existence of a such
form of $\eta $ cannot be assured. However, the example in
Subsec.4.2 suggests that the existence of operators of the form
$\eta =\xi \otimes \zeta $ could  be ensured in case of
$PT$-symmetric Hamiltonians.

A complete characterization of the subclass of quasi-Hermitian Hamiltonians
admitting $\eta $ operators of the form $\eta =\xi \otimes \zeta $ and the
generalization of these results to multipartite quantum systems will be
considered in a forthcoming paper \cite{I}.

We hope that the present developments on alternative descriptions
for quasi-Hermitian dynamics associated with bipartite compound
quantum systems could be also useful, as a preliminary step, in
order to study the entanglement in the context of formulations of
quantum mechanics with non-Hermitian operators and to obtain a
classification of (positive) dynamical maps in the space of
quasi-Hermitian density matrices.

\bigskip

\textbf{Acknowledgements }The authors wishes to thank professor H. F. Jones
and one of the anonymous referees for their suggestions about an erroneous
example in a previous version of the present paper.

\end{document}